\documentclass[twocolumn,epjc3]{svjour3}          

\RequirePackage[T1]{fontenc}

\smartqed  

\RequirePackage[dvipdfmx]{graphicx}
\RequirePackage{mathptmx}      
\RequirePackage{flushend}
\RequirePackage[numbers,sort&compress]{natbib}
\RequirePackage[colorlinks,citecolor=blue,urlcolor=blue,linkcolor=blue]{hyperref}
\usepackage{amsmath,amssymb,tabularx,mathrsfs,bm,braket,ascmac}
\journalname{Eur. Phys. J. A}

\begin{document}

\title{A review of quarkonia under strong magnetic fields}

\author{Sachio Iwasaki\thanksref{e1,addr1}
        \and
        Makoto Oka\thanksref{e2,addr2,addr3}
        \and
        Kei Suzuki\thanksref{e3,addr2} 
}

\thankstext{e1}{e-mail: sutch.iwasaki@th.phys.titech.ac.jp}
\thankstext{e2}{e-mail: oka@post.j-parc.jp}
\thankstext{e3}{e-mail: k.suzuki.2010@th.phys.titech.ac.jp}

\institute{Department of Physics, Tokyo Institute of Technology, Meguro, Tokyo, 152-8551, Japan\label{addr1}
          \and
          Advanced Science Research Center, Japan Atomic Energy Agency, Tokai, Ibaraki, 319-1195, Japan\label{addr2}
          \and
          Nishina Center for Accelerator-Based Science, RIKEN, Wako 351-0198, Japan\label{addr3}
}

\date{Received: 31 March 2021 / Accepted: 20 June 2021}

\maketitle

\begin{abstract}
We review the properties of quarkonia under strong magnetic fields.
The main phenomena are (i) mixing between different spin eigenstates, (ii) quark Landau levels and deformation of wave function, (iii) modification of $\bar{Q}Q$ potential, and (iv) the motional Stark effect.
For theoretical approaches, we review (i) constituent quark models, (ii) effective Lagrangians, (iii) QCD sum rules, and (iv) holographic approaches.
\end{abstract}

\section{Introduction} \label{Sec_1}
In quantum chromodynamics (QCD), the fundamental theory of quarks and gluons, quarks are classified into the six types of flavors: up, down, strange, charm, bottom, and top.
Among them, the charm and bottom quarks are called ``heavy quarks," and hadrons with heavy quarks are sometimes called ``heavy hadrons" such as heavy mesons and heavy baryons.
Quarkonia are bound states composed of a heavy quark and its antiquark.
Since heavy quarks have a heavy mass more than $1$ GeV, heavy hadrons including quarkonia can be produced in high-energy collisions between particles by using particle accelerators (see Ref.~\cite{Brambilla:2010cs,Andronic:2015wma} for comprehensive reviews).
In particular, collisions between heavy nuclei, such as Cu, Au, and Pb, accelerated by the Large Hadron Collider (LHC) and the Relativistic Heavy Ion Collider (RHIC) are called heavy-ion collisions (HICs).
Such experiments are useful to investigate the high-temperature phase of QCD, namely the quark gluon plasma (QGP), where quarks and gluons with color degrees freedom are liberated from the color confinement.

In HICs, when a collision between two nuclei is non-central (or peripheral), a global magnetic field can be produced via the Li\'enard-Wiechert potentials of two moving charged nuclei.
The magnitude of magnetic field depends on its collision energy and impact parameter, and an example in the LHC is estimated to be $|eB|\sim 50 m_\pi^2\sim 1\ \mathrm{GeV}^2\sim 10^{19}\ \mathrm{Gauss}$ ($m_\pi$ is the pion mass) by analytic considerations or numerical simulations \cite{Rafelski:1975rf,Kharzeev:2007jp,Skokov:2009qp,Voronyuk:2011jd,Ou:2011fm,Bzdak:2011yy,Deng:2012pc,Bloczynski:2012en,Bloczynski:2013mca,Deng:2014uja,Huang:2015oca,Zhao:2017rpf,Zhao:2019crj,Cheng:2019qsn}.
Such strong magnetic fields in HICs have attracted much interest, particularly in the viewpoint of chiral transport phenomena and magnetic response of nonperturbative QCD.

In this review, we focus on phenomena in quarkonia under a magnetic field.
The essence of qurkonium properties in a magnetic field can be well known within the constituent quark model with numerical approaches~\cite{Alford:2013jva,Bonati:2015dka,Suzuki:2016kcs,Yoshida:2016xgm,Iwasaki:2018pby}.
Although this model is a simplified model with constituent quark degrees of freedom, it can implement various quantum phenomena induced by a magnetic field.
Some of their properties are confirmed also by QCD sum rules~\cite{Cho:2014exa,Cho:2014loa} and an effective Lagrangian \cite{Cho:2014exa,Cho:2014loa,Yoshida:2016xgm,Mishra:2020kts}.
In particular, there are some characteristic phenomena: (i) the mixing between spin-singlet and spin-triplet eigenstates~\cite{Yang:2011cz,Alford:2013jva,Guo:2015nsa,Bonati:2015dka,Suzuki:2016kcs,Yoshida:2016xgm,Suzuki:2016fof,Dutta:2017pya,Hoelck:2017dby,Iwasaki:2018czv,Iwasaki:2018pby,Chen:2020xsr,Cho:2014exa,Cho:2014loa,Mishra:2020kts,Iwasaki:2021kms,Iwasaki:phd}, which originates from the Zeeman coupling of heavy quarks, (ii) the Landau levels of heavy quarks (or squeezing of spatial wave function), (iii) anisotropic (or modified) confinement potential~\cite{Miransky:2002rp,Chernodub:2014uua,Andreichikov:2012xe,Bonati:2014ksa,Rougemont:2014efa,Simonov:2015yka,Bonati:2015dka,Bonati:2016kxj,Bonati:2017uvz,Hasan:2017fmf,Singh:2017nfa,Hasan:2018kvx,Bagchi:2018mdi,Bonati:2018uwh,Khan:2020gky,Hasan:2020iwa,Zhou:2020ssi}, and (iv) the motional Stark effect (or Lorentz ionization) in moving quarkonia~\cite{Marasinghe:2011bt,Tuchin:2013ie,Alford:2013jva,Bonati:2015dka,Guo:2015nsa,Chen:2020xsr}.
For other phenomenological studies, see Refs.~\cite{Tuchin:2011cg,Machado:2013rta,Dudal:2014jfa,Sadofyev:2015hxa,Braga:2018zlu,Braga:2019yeh,Braga:2020hhs}.
For a part of review papers, see Sec.~IV in Ref.~\cite{Hattori:2016emy} and Sec.~V in Ref.~\cite{Zhao:2020jqu}.

This review paper is organized as follows:
In Sec.~\ref{sec:2}, we summarize physical phenomena realized in magnetized quarkonia, which is the main part of this paper.
Next, in Sec.~\ref{sec:3}, we introduce theoretical approaches to study the magnetic properties of quarkonia.
In realistic HIC experiments, not only the effects from a magnetic field but also contributions from various extreme environments should be considered.
Previous works taking into account finite temperature, finite density, and finite vorticity are commented in Secs.~\ref{sec:4}-\ref{sec:6}, respectively.
In addition, as an example of other hadronic systems, we mention the magnetized heavy-light mesons in Sec.~\ref{sec:7}.
Finally, Sec.~\ref{sec:8} is devoted to the other topics and future prospects.
In \ref{App:1}-\ref{App:3}, we briefly review typical mass spectra and wave functions of $S$-wave charmonia, $S$-wave bottomonia, and $P$-wave charmonia, respectively.

\section{Phenomena in magnetized quarkonia} \label{sec:2}
\begin{figure}[tb!]
    \center
    \begin{minipage}[t]{1.0\columnwidth}
        \begin{center}
            \includegraphics[clip, width=1.0\columnwidth]{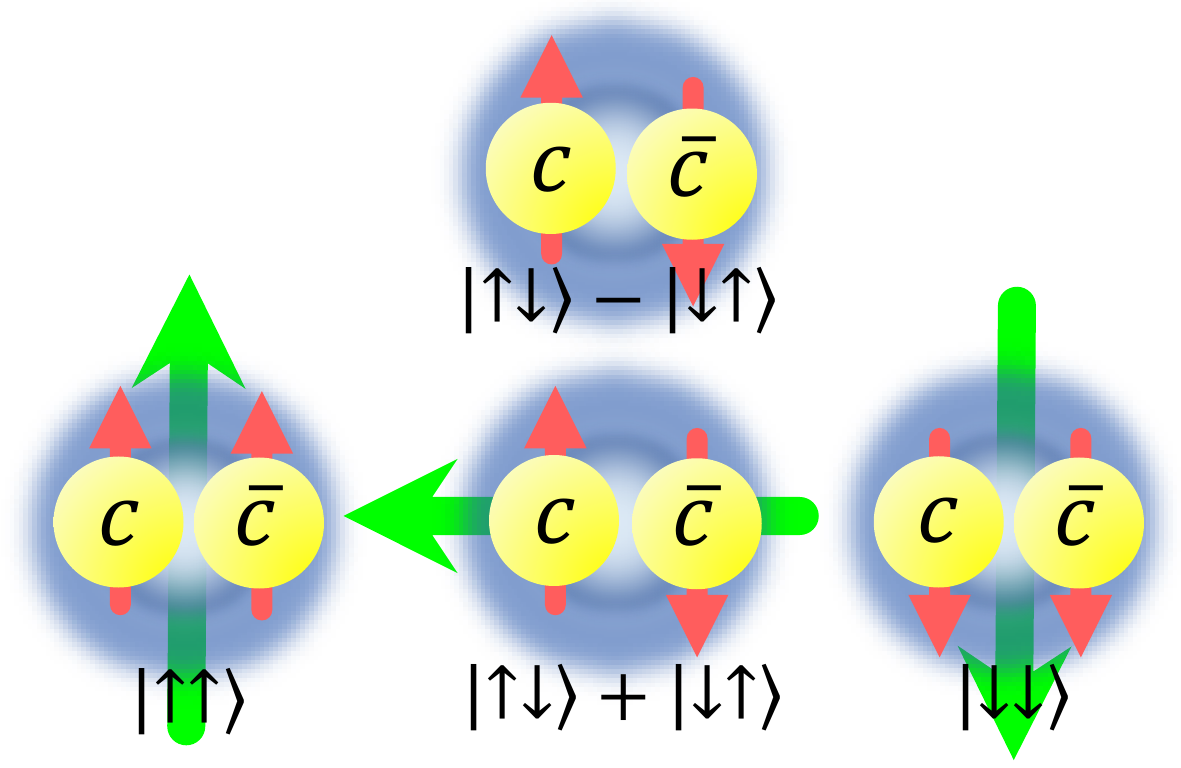}
        \end{center}
    \end{minipage}%
    \caption{Schematic picture of spin eigenstates of $S$-wave charmonia: $\eta_c$ ($\ket{\uparrow \downarrow} - \ket{\downarrow \uparrow}$) and $J/\psi$ ($\ket{\uparrow \uparrow}$, $\ket{\uparrow \downarrow} + \ket{\downarrow \uparrow}$, and $\ket{\downarrow \downarrow}$).}
\label{fig:charmonia}
\end{figure}

In this section, we review the phenomena of quarkonia in a magnetic field.
First, we summarize the notation for the $S$-wave charmonia.
The spin structures of the $S$-wave charmonia at zero magnetic field are characterized by the combination of internal quark spins, and there are the spin singlet (or pseudoscalar) $\eta_c$ and spin triplet (or vector) $J/\psi$ states, as shown in Fig.~\ref{fig:charmonia}.
For these states, we apply the following notations:
\begin{align}
&\ket{\eta_c}:& \frac{1 }{\sqrt{2}} (\ket{\uparrow \downarrow} - \ket{\downarrow \uparrow}) &\equiv \ket{00} \nonumber\\
&\ket{J/\psi,S_z=+1}:& \ket{\uparrow \uparrow} &\equiv \ket{1+1} \nonumber\\
&\ket{J/\psi,S_z=0}:& \frac{1 }{\sqrt{2}} (\ket{\uparrow \downarrow} + \ket{\downarrow \uparrow}) &\equiv \ket{10} \nonumber\\
&\ket{J/\psi,S_z=-1}:& \ket{\downarrow \downarrow} &\equiv \ket{1-1},  \nonumber
\end{align}
where the notation $\ket{SS_z}$ in the right is characterized by the total spin $S$ and its $z$ component $S_z$.
For vector states, $\ket{10}$ and $\ket{1 \pm 1}$ are sometimes called ``longitudinal" and ``transverse" components.
Note that, for bottomonia, the corresponding $S$-wave particles are $\eta_b$ and $\Upsilon$.

\subsection{Mixing between spin eigenstates}
\begin{figure}[tb!]
    \center
    \begin{minipage}[t]{1.0\columnwidth}
        \begin{center}
            \includegraphics[clip, width=1.0\columnwidth]{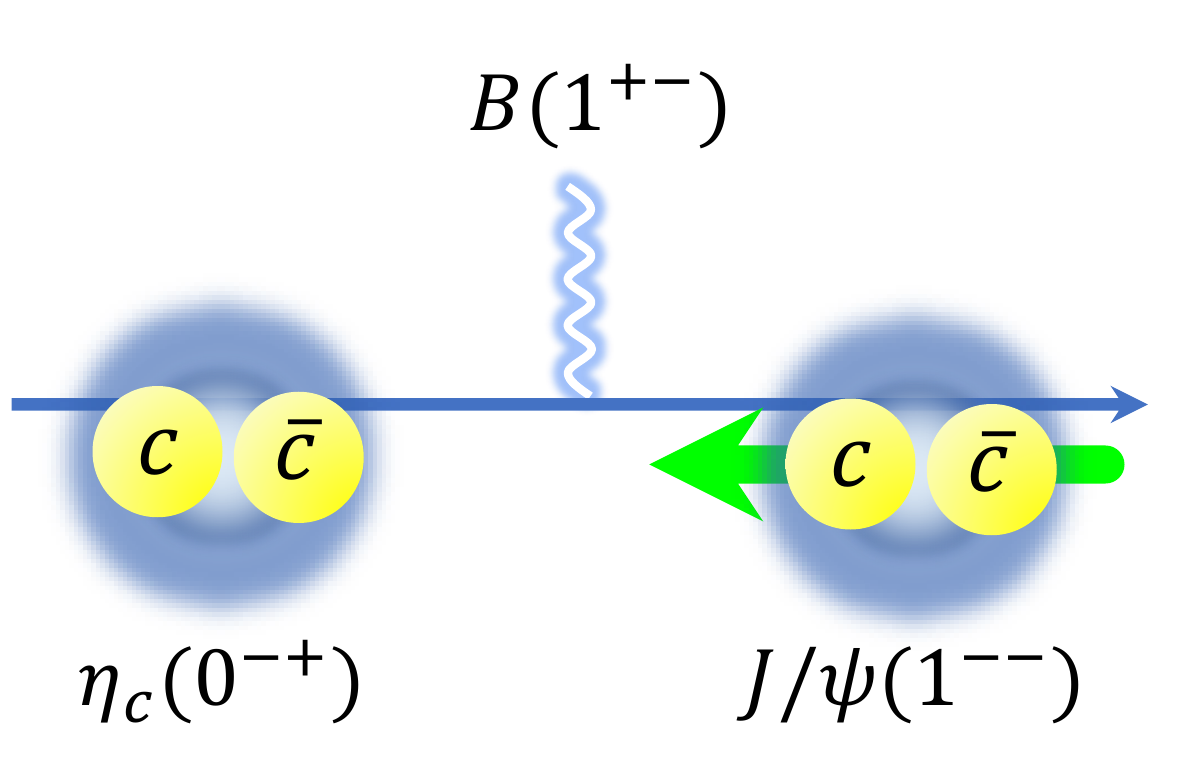}
        \end{center}
    \end{minipage}%
    \caption{Schematic picture of mixing between the spin singlet $\eta_c$ and spin triplet $J/\psi$ in a magnetic field.
The quantum numbers are labeled by $J^{PC}$, where $J$, $P$ and $C$ are the total angular momentum, parity, and charge conjugate, respectively.}
\label{fig:mixing}
\end{figure}

The quantum number of an external magnetic field is $J^{PC}=1^{+-}$, and the magenetic field induces a mixing between a pseudoscalar particle with $J^{PC}=0^{-+}$ and a vector particle with $J^{PC}=1^{--}$ (see Fig.~\ref{fig:mixing}).
Hence, for quarkonia, one can expect mixing such as $\eta_c$-$J/\psi$ and $\eta_b$-$\Upsilon$.
Such mixing phenomena can be described by Hamiltonian for the spin degrees of freedom. 
The coupling term between quark magnetic moment and external magnetic field $\bm{B}$ is written as
\begin{align}
H_\mathrm{m.m.} = -\sum_{i=1}^2 \bm{\mu}_i \cdot \bm{B}, \label{eq:Hmm}
\end{align}
where the indices $i=1$ and $2$ correspond to a heavy quark and its antiquark, respectively, and $\bm{\mu}_i = g q_i \bm{S}_i/2m_i$ is the magnetic moment of the $i$ th particle.
$g$, $q_i$, $\bm{S}_i$, and $m_i$ are the Land\'e $g$-factor, electric charge, spin operator, and mass of the $i$ th particle, respectively.
The Hamiltonian~(\ref{eq:Hmm}) induces a mixing between the spin-singlet and the $S_z=0$ component of the spin-triplet eigenstates of quarkonia:
\begin{align}
\bra{10} H_\mathrm{m.m.} \ket{00} &= -\frac{gB}{4} \left( \frac{q_1}{m_1} - \frac{q_2}{m_2} \right) \label{eq:spin_mix_1} \\
&\to -\frac{gqB}{4} \left( \frac{1}{m_1} + \frac{1}{m_2} \right) \neq 0, \label{eq:spin_mix_2} \\
\bra{00} H_\mathrm{m.m.} \ket{00} &= \bra{10} H_\mathrm{m.m.} \ket{10} =0, 
\end{align}
where, ``$\to$" indicates that we impose the condition of neutral meson ($q_1=-q_2=q$), and we also used the eigenvalues from the spin operator and the two-body spin eigenstates,
\begin{align}
\bm{S}_1 \cdot \bm{B} \left( \ket{\uparrow \downarrow} \pm \ket{\downarrow \uparrow} \right)&=  \frac{B}{2} \left[ \ket{\uparrow \downarrow} \mp  \ket{\downarrow \uparrow} \right], \nonumber\\
\bm{S}_2 \cdot \bm{B} \left(\ket{\uparrow \downarrow} \pm \ket{\downarrow \uparrow}\right) &= \frac{B}{2} \left[- \ket{\uparrow \downarrow} \pm  \ket{\downarrow \uparrow} \right]. \nonumber
\end{align}

From the nonzero matrix element~(\ref{eq:spin_mix_2}), we can see that the spin quantum number of hadrons is no longer good due to the presence of a magnetic field, whereas $S_z$ is still a good quantum number.
Such mixing effect leads to the mass shifts of hadrons by the level repulsion: as the magnetic field increases and the mixing becomes stronger, the mass of the lower state decreases while that of the higher state increases (for an example, see the figures of mass spectra in \ref{App:1}-\ref{App:3}).

On the other hand, when the Hamiltonian~(\ref{eq:Hmm}) acts on the $S_z=\pm1$ component of the spin-triplet eigenstates, its eigenvalue is zero:
\begin{align}
H_\mathrm{m.m.} \ket{1\pm 1} &= \mp \frac{gB}{4} \left( \frac{q_1}{m_1} + \frac{q_2}{m_2} \right) \ket{1\pm 1} \label{eq:zeeman_1} \\
&\to \mp \frac{gqB}{4} \left( \frac{1}{m_1} - \frac{1}{m_2} \right) \ket{1\pm 1} = 0.
\end{align}
Since $m_1=m_2$ for quarkonia, the $S_z=\pm1$ components are not affected by the Hamiltonian~(\ref{eq:Hmm}).
This is because the masses of the two particles in quarkonia are identical, and their electric charges are opposite.
If $m_1 \neq m_2$ and/or $q_1 \neq -q_2$, then the energies of the $S_z=\pm1$ components of the spin-triplet hadrons are split, which is nothing but the anomalous Zeeman splitting via the nonzero magnetic moments of hadrons.
This situation is realized in not only charged mesons such as $D^\pm$ mesons but also neutral mesons such as $D^0$ and $\bar{D}^0$ mesons (see Sec.~\ref{sec:7}).

The first comprehensive understanding of this effect was from the constituent quark model \cite{Alford:2013jva}, which was motivated by the $J/\psi$ production from $\eta_c$ in a magnetic field in Ref.~\cite{Yang:2011cz} and the spin mixing in $D$ mesons in a magnetic field in Ref.~\cite{Machado:2013rta}.
In Ref.~\cite{Alford:2013jva}, the spin mixing for the ground states of charmonia and bottmonia was examined.
After that, a similar tendency was obtained from other works with the same quark model \cite{Bonati:2015dka,Suzuki:2016kcs,Yoshida:2016xgm} and from QCD sum rules and effective Lagrangians \cite{Cho:2014exa,Cho:2014loa}.
 
The $P$-wave quarkonia in a magnetic field were studied within the constituent quark model in Refs.~\cite{Bonati:2015dka,Iwasaki:2018pby}.
The mixing structure in $P$-wave states is more complicated than that of $S$-waves.
For example, in $P$-wave charmonia, some components of $\chi_{c0}$ with $J^{PC}=0^{++}$, $\chi_{c1}$ with $J^{PC}=1^{++}$, and $\chi_{c2}$ with $J^{PC}=2^{++}$ are mixed with $h_c$ with $J^{PC}=1^{+-}$.
Since the $P$-wave quarkonia have nonzero orbital angular momentum, which is characterized by the LS and tensor couplings, the behavior of the spatial wave function is quite different from that of $S$-waves.
In particular, in Ref.~\cite{Iwasaki:2018pby}, the authors pointed out the {\it hadronic Paschen-Back effect} (HPBE) (see also \ref{App:3}), which is analogous to the phenomenon observed in hydrogen atoms~\cite{Paschen:1921}.
The HPBE is induced by magnetic fields comparable with the LS splitting of the hadron masses, so that the wave functions of $P$-wave quarkonia can be deformed even in relatively smaller magnetic fields than those of the $S$-wave quarkonia.
One of the resultant phenomena is the anisotropic photon emission in radiative decays \cite{Iwasaki:2018pby,Iwasaki:2018czv} such as $\chi_{c0,c1,c2} \to J/\psi \gamma$.
As other phenomena relating to the HPBE, the anisotropy in strong decays and quarkonium production will be interesting.

\subsection{Quark Landau levels}

\begin{figure}[tb!]
    \center
    \begin{minipage}[t]{1.0\columnwidth}
        \begin{center}
            \includegraphics[clip, width=1.0\columnwidth]{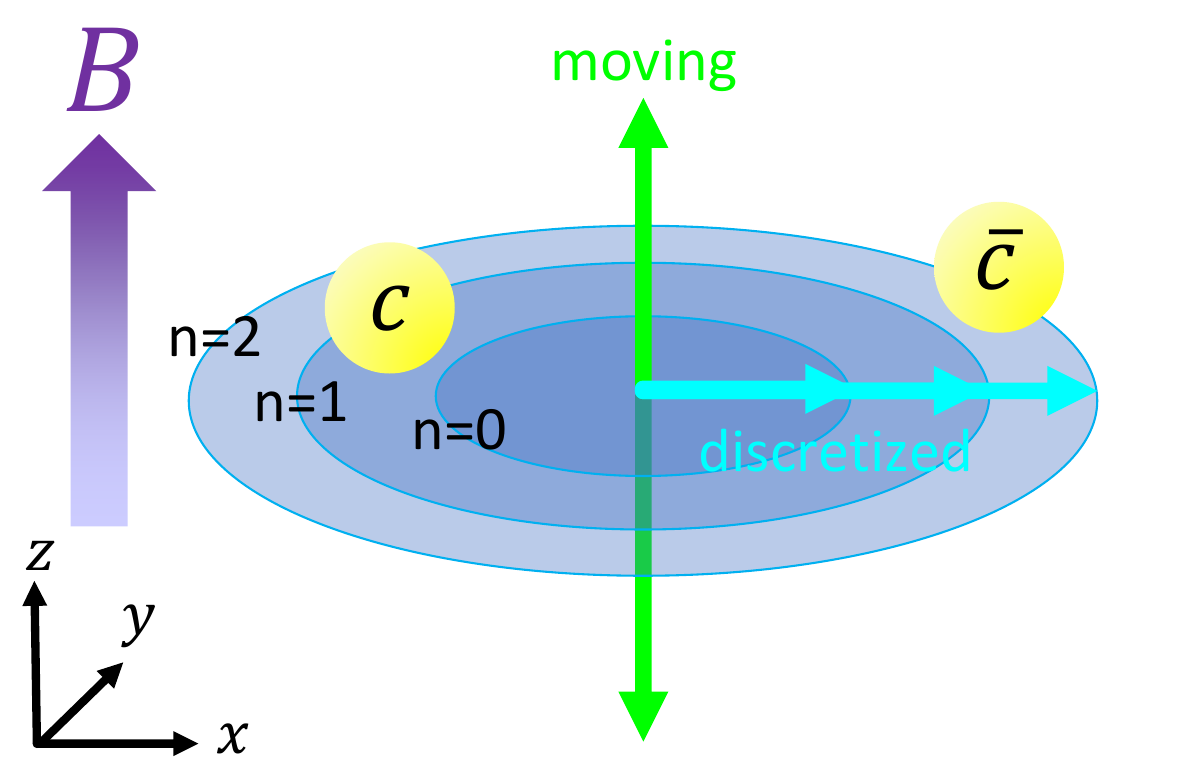}
        \end{center}
    \end{minipage}%
    \caption{Schematic picture of Landau levels of charm quarks.
While the momentum for the $\rho=\sqrt{x^2+y^2}$ direction are discretized, the momentum for the $z$ direction is not affected when a magnetic field parallel to the $z$ axis is applied.}
\label{fig:quarkLLs}
\end{figure}

In general, for a charged particle in a magnetic field, the transverse momenta, namely the momenta perpendicular to the magnetic field, are discretized and each level is characterized by an integer label.
Such a discretized levels are called Landau levels (See Fig.~\ref{fig:quarkLLs}).
In nonrelativistic quark model, the Landau levels of nonrelativistic constituent quarks are induced by the coupling between original kinetic term and a magnetic field via the vector potential $\bm{A}_i \equiv \bm{A}(\bm{r}_i)$, where $\bm{r}_i$ is the coordinate of $i$ th particle:
\begin{align}
H_\mathrm{kin}^\mathrm{2body} = \sum_{i=1}^{2} \frac{1}{2m_i} (\bm{p}_i -q_i \bm{A}_i)^2,
\end{align}
where $m_i$, $\bm{p}_i$, and $q_i$ are the mass, momentum, and electric charge of $i$ th particle, respectively.
The Landau levels of $i$ th particle are induced by the quadratic term proportional to $(-q_i \bm{A}_i)^2$.
After the original two-body Hamiltonian is reduced to the relative Hamiltonian, we obtain the term corresponding to the Landau levels for relative motions:
\begin{align}
H_\mathrm{LLs} = \frac{q^2}{8m}(\bm{B} \times \bm{r})^2 = \frac{q^2B^2}{8m} \rho^2, \label{eq:LLs}
\end{align}
where we used $\bm{B} =(0,0,B)$ and the symmetric gauge $\bm{A}_i = \frac{1}{2} \bm{B} \times \bm{r}_i$.
$q=|q_1|=|q_2|$, $m=m_1/2=m_2/2$ is the reduced mass, $\bm{r}=\bm{r}_1-\bm{r}_2$ is the relative coordinate, and $\rho =\sqrt{x^2+y^2}$ is the transverse component of the cylindrical coordinate.
Thus, the quark Lanudau levels show an enhancement of the energy, which is independent of the signs of electric charges of quarks.
The dependence of $\rho^2$ for the symmetric gauge is regarded as the two-dimensional harmonic-oscillator potential on the $\rho$ plane.
Such a spatially anisotropic potential leads to deformation of the wave functions of quarkonia~\cite{Suzuki:2016kcs,Yoshida:2016xgm} (see the figures in \ref{App:1}-\ref{App:3}).
Wave function deformation under magnetic fields can be measured by lattice QCD simulations as in Ref.~\cite{Hattori:2019ijy}.

Note that the term proportional to $(\bm{B} \times \bm{r}) \cdot \bm{p} = \bm{B} \cdot (\bm{r} \times \bm{p})$, namely the coupling of the magnetic field and an orbital angular momentum, leads to the normal Zeeman effect.
This term is proportional to $\frac{q_1}{m_1} + \frac{q_2}{m_2}$, which is similar to the anomalous Zeeman effect~(\ref{eq:zeeman_1}), so that there is no contribution to quarkonia.

\subsection{Deformation of $\bar{Q}Q$ potential}

\begin{figure}[tb!]
    \center
    \begin{minipage}[t]{1.0\columnwidth}
        \begin{center}
            \includegraphics[clip, width=1.0\columnwidth]{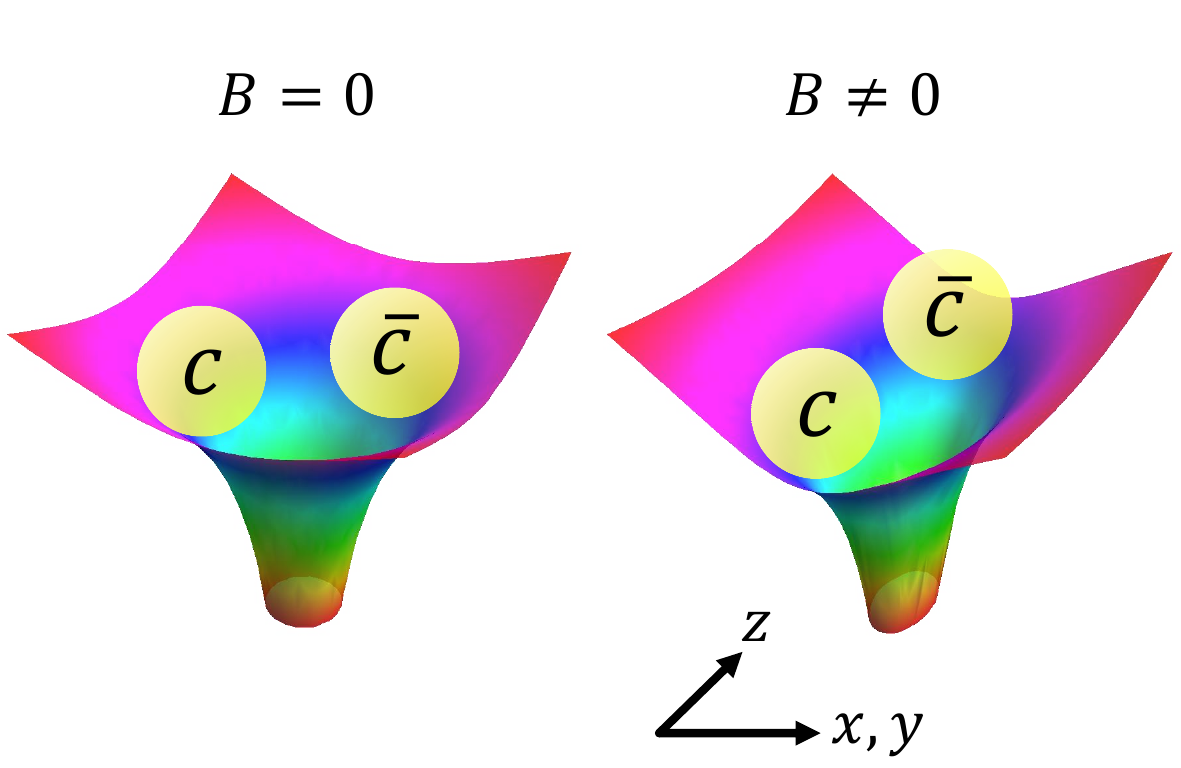}
        \end{center}
    \end{minipage}%
    \caption{Schematic picture of $\bar{Q}Q$ potential anisotropically deformed by a magnetic field.
Left: $\bar{Q}Q$ potential is spatially isotropic at vanishing magnetic field.
Right: $\bar{Q}Q$ potential becomes anisotropic at nonzero magnetic field.}
\label{fig:potential}
\end{figure}

Many hadrons including quarkonia can be described in nonrelativistic quantum mechanics, where hadrons are regarded as bound states confined by the $\bar{Q}Q$ potential.
The main part of the $\bar{Q}Q$ potential is considered to be composed of the linear and Coulomb potentials:
\begin{align}
H_\mathrm{pot.} &= V(\bm{r}) \nonumber\\
&= \sigma r - \frac{4}{3} \frac{\alpha_s}{r} + \cdots,
\end{align}
where  $\bm{r}=\bm{r}_1-\bm{r}_2$ is the relative coordinate, and $r$ is the radial component for the spherically-symmetric potential.
$\sigma$ and $\frac{4}{3} \alpha$ are the coefficients of the linear and Coulomb potentials, respectively.
``$\cdots$" includes the higher-order terms such as the LS coupling and tensor coupling, which can be included as relativistic corrections based on QCD.
In zero magnetic field, these potentials are spatially isotropic which reflects the spherical symmetry of the vacuum (see the left of Fig.~\ref{fig:potential} for a typical potential).

In a nonzero magnetic field, these potentials can be modified, and not only the coefficients, $\sigma$ and $\frac{4}{3} \alpha$, but also the functional form of the potential depend on the magnetic field.
In particular, the spatial form of the potential can be anisotropic due to the breaking of the spherical symmetry (see the right of Fig.~\ref{fig:potential}).
Such an anisotropy has been predicted from effective models \cite{Miransky:2002rp,Chernodub:2014uua,Andreichikov:2012xe,Simonov:2015yka} and holographic approaches~\cite{Rougemont:2014efa,Zhou:2020ssi}.
In fact, the anisotropy of the static potential for the linear and Coulomb parts was measured by $N_f=2+1$ lattice QCD simulations at zero temperature \cite{Bonati:2014ksa} and at finite temperature \cite{Bonati:2016kxj}. 
The influence in the constituent quark model by inputting lattice results was studied in Ref.~\cite{Bonati:2015dka}.
As other observables in lattice QCD simulations, the anisotropy was observed also in electric and magnetic screening masses defined by Polyakov loop correlators~\cite{Bonati:2017uvz} and flux tube profile defined by the combination of the Wilson loop and the plaquette operators~\cite{Bonati:2018uwh}.

\subsection{Motional Stark effect}

\begin{figure}[tb!]
    \center
    \begin{minipage}[t]{1.0\columnwidth}
        \begin{center}
            \includegraphics[clip, width=1.0\columnwidth]{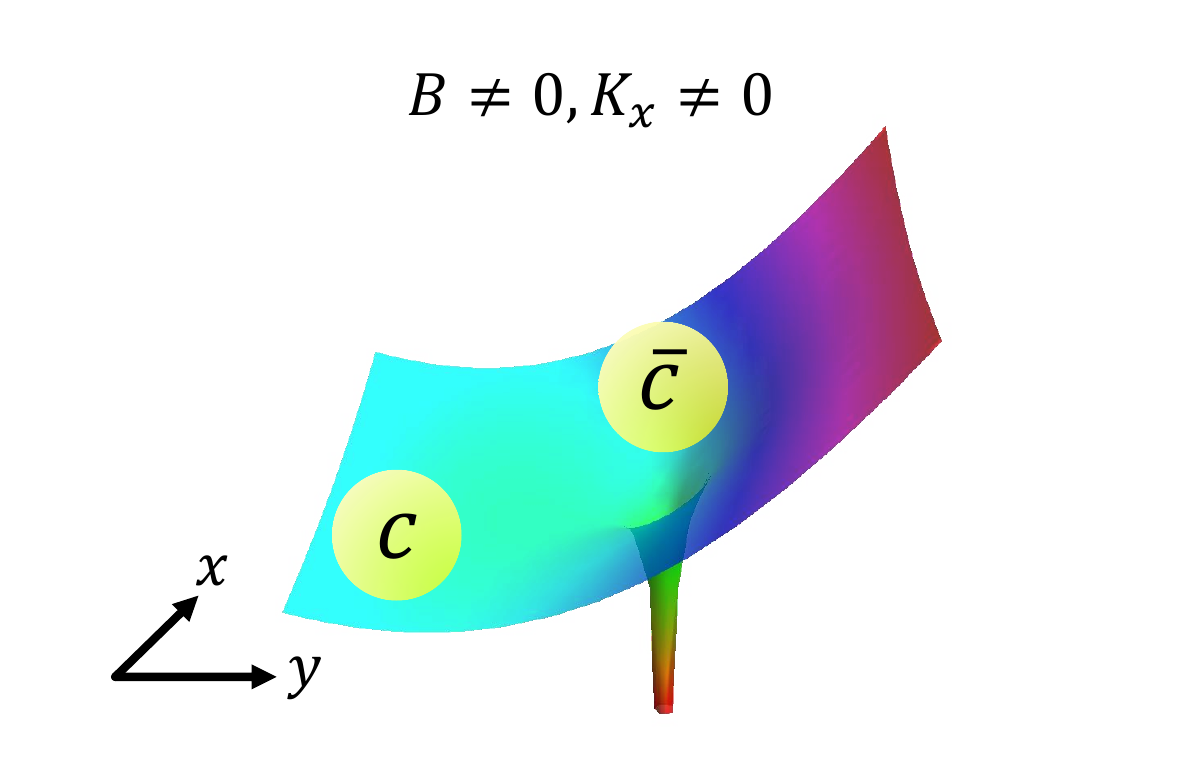}
        \end{center}
    \end{minipage}%
    \caption{Schematic picture of Lorentz ionization by the motional Stark effect of charm quarks.}
\label{fig:MSE}
\end{figure}

In the laboratory frame, a charged particle with a velocity $\bm{v}$ in a magnetic field $\bm{B}$ is affected by the Lorentz force $\bm{v} \times \bm{B}$.
Equivalently, in the comoving frame, this particle is affected by an effective electric field $\bm{E} = \bm{v} \times \bm{B}$.
This effect is called the {\it motional Stark effect}~\cite{Lamb:1952zza}.

The original momentum $\bm{p}$ of a charged particle under a magnetic field is not conserved, but we can define a new conserved quantity~\cite{Johnson:1949}, $\bm{p} + q \bm{A}$, which is the so-called {\it pseudomomentum} for a single particle.
Similarly, in two-body systems composed of two charged particles in a magnetic field, the momentum of the center of mass motion is not conserved.
Instead, we can define the pseudomomentum for the two-body system~\cite{Gor'kov:1968,Carter:1969aj,Avron:1978}:
\begin{equation}
\bm{K} = \sum_{i=1}^2 \left[ \bm{p}_i + q_i \bm{A}_i \right].
\end{equation}
This operator commutes with the original Hamiltonian.
If one assumes that the pseudomomentum $\bm{K}=(K_x,K_y,K_z)$ is zero, the Hamiltonian of the relative motion becomes a simple form, where the terms depending on the magnetic field are only the magnetic-moment term (\ref{eq:Hmm}) and Landau levels term (\ref{eq:LLs}).
When the pseudomomentum is nonzero, the $\bm{K}$ dependent terms appear, which characterize the interplay between the center of mass motion and relative motion:
\begin{align}
H_\mathrm{pse.} &= \frac{\bm{K}^2}{2M}-\frac{q}{M} (\bm{K} \times \bm{B}) \cdot \bm{r} \nonumber\\
&= \frac{\bm{K}^2}{2M} + \frac{qB}{4m} (K_x y -K_y x),
\end{align}
where we used $\bm{B} =(0,0,B)$ and the symmetric gauge $\bm{A}_i = \frac{1}{2} \bm{B} \times \bm{r}_i$.
$q=q_1=-q_2$, $M=m_1+m_2$, and $m=m_1/2=m_2/2$.
The first term corresponds to just an energy shift, and the second term is a pseudomomentum-dependent potential.
As $K_x$ increases, the transverse direction of the potential is deformed as shown in Fig.~\ref{fig:MSE}.
If $K_x$ is large enough, a local minimum in the potential appears, and the wave function is delocalized, which is the Lorentz ionization.
Strictly speaking, as long as the quarks are confined by the confinement potential, qurkonia are not ionized.

The Lorentz ionization for charmonia was first predicted in Ref.~\cite{Marasinghe:2011bt} (also see Ref.~\cite{Tuchin:2013ie} for a review).
After that, this effect was reinterpreted within the relativistic Hamiltonian of the constituent quark model \cite{Alford:2013jva}.
For other discussions, see Refs.~\cite{Marasinghe:2011bt,Tuchin:2013ie,Alford:2013jva,Bonati:2015dka,Guo:2015nsa,Chen:2020xsr}.

\section{Theoretical approaches for magnetized quarkonia} \label{sec:3}
\subsection{Constituent quark model}
The constituent quark model~\cite{Eichten:1974af} is known to be an approach to study quantum-mechanical bound states by solving the Schr\"odinger equation with constituent quark degrees of freedom and an effective potential between quarks.
Although this approach is a very simplified model, compared to QCD, it can well reproduce the various observed properties of hadrons, such as the mass spectrum and decay properties.
In the potential models for Coulomb systems such as the hydrogen atom and the positronium, the effects of a (strong) magnetic field has been studied for a long time.
On the other hand, the quark models for quarkonia in a magnetic field were recently investigated~\cite{Alford:2013jva,Bonati:2015dka,Suzuki:2016kcs,Yoshida:2016xgm,Iwasaki:2018pby}.
One of the characteristics of such a model is the presence of the linear (or confining) potential which is, of course, implemented to reproduce the color confinement realized in the low-energy regime of QCD.
In this sense, this system is qualitatively different from the well-known Coulomb systems.

The first application to $S$-wave charmonia ($\eta_c$ and $J/\psi$) and $S$-wave bottomonia ($\eta_b$ and $\Upsilon$) was done in Ref.~\cite{Alford:2013jva}, where the spin mixing and behaviors at finite pseudomomentum were also taken into account.
After that, in Ref.~\cite{Bonati:2015dka}, the authors investigated the corrections from the anisotropy of the potential measured from lattice QCD simulations~\cite{Bonati:2014ksa}.
In Refs.~\cite{Suzuki:2016kcs,Yoshida:2016xgm}, the authors studied the excited states of $S$-wave charmonia (such as $\eta_c^\prime$ and $\psi^\prime$) and $S$-wave bottomonia as well as neutral $D$, $B$, and $B_s$ mesons.
For the $P$-wave charmonia ($h_c$, $\chi_{c0}$, $\chi_{c1}$, and $\chi_{c2}$), see Refs.~\cite{Bonati:2015dka,Iwasaki:2018pby}.

In general, the application of the constituent quark model is not limited to quarkonia and heavy-flavor hadrons (as long as the constituent-quark picture is valid).
Therefore, similar analyses with a magnetic field can be applied to mesons \cite{Simonov:2012if,Simonov:2013jpa,Andreichikov:2012xe,Andreichikov:2013zba,Orlovsky:2013gha,Taya:2014nha,Andreichikov:2016ayj,Kojo:2021gvm} and baryons \cite{Andreichikov:2013pga,Taya:2014nha} composed of light (up, down, or strange) quarks.

\subsection{Effective Lagrangian}
Effective Lagrangians with hadronic degrees of freedom are other approaches to describe the properties of hadrons in a magnetic field.
In particular, in a weak-magnetic-field region, the mass shift from the spin mixing would be dominant. 
The spin mixing is originally understood via the magnetic moment of the quark degrees of freedom, but their properties can be described also by an effective Lagrangian.

In Refs.~\cite{Cho:2014exa,Cho:2014loa}, the authors considered the following Lagrangian with a massive pseudoscalar field $P$ and a massive vector field $V^\mu$:
\begin{align}
\mathcal{L}_{\mathrm{P}} &= \frac{1}{2} \partial_\mu P \partial^\mu P - \frac{1}{2} m_\mathrm{P}^2 P^2, \\
\mathcal{L}_{\mathrm{V}} &= \frac{1}{2} \partial_\mu V_\nu \partial^\mu V^\nu - \frac{1}{2} m_\mathrm{V}^2 V_\nu V^\nu, \\
\mathcal{L}_{\gamma \mathrm{PV}} &= \frac{g_\mathrm{PV}}{m_0} e\tilde{F}_{\mu\nu} (\partial^\mu P) V^\nu,
\end{align}
where $g_\mathrm{PV}$ is the dimensionless coupling constant, and $m_0=(m_\mathrm{P}+m_\mathrm{V})/2$ is the average of masses of the pseudoscalar and vector field at zero magnetic field.
From these Lagrangians, the equations of motion are
\begin{align}
(\partial^2 + m_\mathrm{P}^2)P + \frac{g_\mathrm{PV}}{m_0} e\tilde{F}_{\mu\nu} \partial^\mu V^\nu =0, \\
(\partial^2 + m_\mathrm{V}^2)V_\nu - \frac{g_\mathrm{PV}}{m_0} e\tilde{F}_{\mu\nu} (\partial^\mu P) =0.
\end{align}
Here, we assume that a magnetic field is parallel to to the $z$ direction, and then the survival components of the dual electromagnetic field tensor are $\tilde{F}_{03} = - \tilde{F}_{30} = B$.
When the spatial momenta of quarkonia are zero, $p_\mu =(p_0,0,0,0)$, the survival components are only the temporal derivative $\partial^0 \to -ip^0 =-ip_0$ in momentum space and the third component of the vector field $V^\mu = (V_0, \bm{V}_\perp, V_\parallel)$.
The matrix form of the equations of motion is
\begin{align}
\left(
\begin{array}{ccc}
- p_0^{2} +  m_\mathrm{P}^{2}  && -i \dfrac{g_\mathrm{PV}}{m_0} p_0 eB \\
i \dfrac{g_\mathrm{PV}}{m_0} p_0 eB && - p_0^{2} +  m_\mathrm{V}^2 \\
\end{array}
\right)
\left(
\begin{array}{c}
P \\
V_\parallel \\
\end{array}
\right)
=0. \label{HadronEFT1}
\end{align}
By diagonalizing this matrix, one can get the mass formulas including the mixing effect.

The effective Lagrangian with the mixing including radially excited states such as $2S$ states was constructed in Ref.~\cite{Yoshida:2016xgm}, where the authors considered the $\eta_c$-$J/\psi$-$\eta_c^\prime$-$\psi^\prime$ mixing.
The mixing including the $1D$ state was studied in Ref.~\cite{Mishra:2020kts}.

\subsection{QCD sum rules}
QCD sum rules~\cite{Shifman:1978bx,Shifman:1978by} are powerful approaches to extract the spectral functions of the hadrons from the operator product expansion (OPE) of the hadronic correlation functions (or hadronic correlators) of QCD.
For example, the Borel sum rule is written as
\begin{align}
\mathcal{M}_\mathrm{OPE}^J(M^2) = \int \rho^J(s) e^{-s/M^2} ds.
\end{align}
The right-hand side is the integral of the spectral function $\rho^J(s)$ with a weighting factor $e^{-s/M^2}$ over the energy $\sqrt{s}$, and the spectral function $\rho^J(s)$ includes the properties of hadrons with a quantum number $J$, such as the mass and decay widths.
The left-hand side is called the OPE side.
An OPE is composed of the perturbative part and nonperturbative parts including QCD ``condensates" such as the chiral condensate $\braket{\bar{q}q}$ and gluon condensate $\braket{G_{\mu\nu}^aG^{a\mu\nu}}$.
Thus one can study the nonperturbative aspect of hadronic physical quantities via QCD condensates.
The values of condensates are modified by external environments such as finite temperature, density, and magnetic fields.
Once we know the modified values of condensates, then we can predict the change of the properties of hadrons by QCD sum rules.

In order to study magnetic-field effects by QCD sum rules, one has to calculate the magnetic-field dependence of the OPE side.
As a result, one can extract the magnetic response of a spectral function.
QCD sum rules for $S$-wave charmonia are studied in Refs.~\cite{Cho:2014exa,Cho:2014loa}.
Here, the authors calculated the OPEs of the pseudoscalar and vector correlators with a magnetic field.
For the spectral function from the pseudoscalar correlator, they applied a two-peak structure ansatz corresponding to the original $\eta_c$ and a magnetically induced $J/\psi$, which reflects the mixing between $\eta_c$ and $J/\psi$.
As a result, they concluded a negative mass shift of $\eta_c$ and a positive shift of $J/\psi$ in the weak-field region.
These mass shifts were interpreted as level repulsions induced by the mixing between $\eta_c$ and $J/\psi$.
Note that QCD sum rule approaches with a magnetic field were also applied to neutral and charged pseudoscalar $D$ mesons~\cite{Gubler:2015qok} and charged pseudoscalar $B$ mesons \cite{Machado:2013yaa}.

\subsection{Holographic approach}
Holographic approaches based on the AdS/CFT correspondence~\cite{Maldacena:1997re} are other approaches to nonperturbatively study the magnetic properties of hadrons.
In the first study by Dudal and Mertens~\cite{Dudal:2014jfa}, they considered a soft wall model \cite{Karch:2006pv} extended by a Dirac-Born-Infeld (DBI) action.
Here, they introduced the DBI action in order to probe the internal electric charges of quarkonia, which is different from the usual Maxwell action where $J/\psi$ is treated as the point-like particle. 
From this action, they investigated the magnetic-field dependence of the spectral functions of charmonia at low temperature and its melting behavior at high temperature.
The same approach was applied to the diffusion constant of charm quarks~\cite{Dudal:2014jfa,Dudal:2018rki}, since the low-energy limit of a spectral function is proportional to the diffusion constant.
As another model, in Refs.~\cite{Braga:2018zlu,Braga:2019yeh,Braga:2020hhs}, the magnetic field was introduced via the Einstein-Maxwell action.
They used a generalized soft wall model that can reproduce the decay constants of vector charmonia and bottomonia in vacuum, and studied the spectral function at finite temperature.
In addition, a holographic approach can investigate the $\bar{Q}Q$ potential. 
In Refs.~\cite{Rougemont:2014efa,Zhou:2020ssi}, the authors studied the magnetic response of $\bar{Q}Q$ potential, such as the distance between the $\bar{Q}Q$ pair, free energy of the $\bar{Q}Q$ pair, entropy of the $\bar{Q}Q$ systems at finite temperature, and binding energy of the quarkonium.

\section{Quarkonia in magnetized thermal matter} \label{sec:4}
The properties of quarkonia at finite temperature has attracted much attention for a long time in the context of HIC experiments since high-energy collisions can produce hot QCD matter or QGP.
In particular, a famous phenomenon is the $J/\psi$ suppression~\cite{Matsui:1986dk} or associated mass shifts~\cite{Hashimoto:1986nn} (for reviews, e.g., see Refs.~\cite{Gerschel:1998zi,Vogt:1999cu,Satz:2005hx,Rapp:2008tf,Mocsy:2013syh,Rothkopf:2019ipj}).
Here the dissociation of quarkonia could be described by the deformation or disappearance of the confinement potential by the Debye screening for color charges.

Quarkonia in magnetized thermal matter were investigated by holographic approaches~\cite{Dudal:2014jfa,Dudal:2018rki,Braga:2018zlu,Braga:2019yeh,Braga:2020hhs} as well as constituent quark models based on a potential obtained from perturbative calculations with a thermal gluon propagator~\cite{Hasan:2017fmf,Singh:2017nfa,Hasan:2018kvx,Bagchi:2018mdi,Khan:2020gky,Hasan:2020iwa}.
In particular, the real part of the thermal modification of $\bar{Q}Q$ potential is related to the Debye screening, while the imaginary part also appears.
The origins of the imaginary part are regarded as two distinct mechanisms: (i) the Landau damping~\cite{Laine:2006ns} which is the inelastic scattering between heavy quarks and virtual gluons and (ii) the singlet-to-octet transitions~\cite{Brambilla:2008cx,Brambilla:2011sg} which is an absorption process of gluons into color-singlet quarkonia.
The magnetic response of the static $\bar{Q}Q$ potential at finite temperature was studied by lattice QCD simulations~\cite{Bonati:2016kxj,Bonati:2017uvz,Bonati:2018uwh} and holographic approaches \cite{Rougemont:2014efa,Zhou:2020ssi}.

\section{Quarkonia in magnetized dense matter} \label{sec:5}
In relatively low-energy HICs, not only thermal matter but also a finite-density environment can be created, which will be useful to clarify the high-density region of the QCD phase diagram.
The ground state in the high-density region of QCD is quark matter, while the ground state in the relatively low-density region is the nuclear matter composed of protons and neutrons.
When the numbers of protons and neutrons are imbalanced, as realized in neutron rich nuclei and neutron stars, the nuclear matter is called isospin asymmetric nuclear matter.
The properties of quarkonia in nuclear matter are little known so far (see Ref.~\cite{Hosaka:2016ypm} for a review).
In the viewpoint of QCD sum rules, their masses are expected to be affected by modification of the gluon condensates~\cite{Klingl:1998sr,Kim:2000kj}, where mass shifts of a few MeV for $\eta_c$ and $J/\psi$ were estimated.

Even in low-energy HICs, peripheral collision can produce a magnetic field.
Under a magnetic field, the nuclear matter is magnetized, and its property including gluon condensates should be modified.
Quarkonia in isospin symmetric or asymmetric magnetized nuclear matter were studied by effective models~\cite{Jahan:2018jql,Jahan:2018yud,Jahan:2021krw} and QCD sum rules~\cite{Kumar:2018ujk,Parui:2018qwx,Kumar:2019tiw,Parui:2021jbu}.
In particular, the studies of in-medium partial decay width, such as charmonia to $D \bar{D}$ and bottomonia to $B \bar{B}$, might be interesting as an observable, as emphasized in Refs.~\cite{Mishra:2018mag,Mishra:2019ixd,Mishra:2019ljw,Kumar:2019axp}.

\section{Quarkonia in magnetized rotating matter} \label{sec:6}
In HIC experiments, peripheral collisions are expected to produce rotating medium with a total angular momentum (or finite vorticity).
In fact, the evidence of a strong vorticity was observed as a global polarization of $\Lambda$ and $\bar{\Lambda}$ hyperons by STAR Collaboration at RHIC~\cite{STAR:2017ckg,Adam:2018ivw}.
The strength of the vorticity was estimated to be $\omega \approx 9\times 10^{21}$ s$^{-1}$~\cite{STAR:2017ckg}.
The observed difference between polarizations of $\Lambda$ and $\bar{\Lambda}$ might be caused by an electromagnetic field.

Thus, the properties of hadrons in finite-vorticity environments are also interesting.
For quarkonia in rotating matter, there is only one study~\cite{Chen:2020xsr} (also see Ref.~\cite{Zhao:2020jqu} for a review).
The vorticity is coupled to the total angular momentum $\bm{J}$ of a hadron, including the spin $\bm{S}$ and orbital angular momenta $\bm{L}$, by the coupling $\bm{J} \cdot \bm{\omega}$, where $\bm{J} = \bm{L} + \bm{S}$ and $\bm{\omega}$ is the vorticity.
Therefore, $\eta_c$ and longitudinal components of $J/\psi$ are not affected by the vorticity, whereas the energies of the transverse components of $J/\psi$ are shifted by the vorticity.
Here, the spin of a hadron is not a good quantum number, but the third component along the vorticity vector is a good quantum number.
The vorticity does not mix spin eigenstates of hadrons, which is distinct from situations in magnetic fields.
The difference between a magnetic field and a vorticity stems from the existence or absence of coupling to electric charges of quarks.
In this sense, we can distinguish a magnetic-field effect and a vorticity effect through observables for quarkonia.

\section{Heavy-light mesons} \label{sec:7}
Heavy-light mesons are composed of one light ($u$, $d$ or $s$) quark and one heavy quark, which is sometimes called open-heavy-flavor mesons denoting a nonzero charm/bottomness quantum number.
Such hadrons are important in the sense that these are related to the two-meson threshold of quarkonia.
As with quarkonia, the spin-singlet and the longitudinal component of spin-triplet are mixed via the coupling of a magnetic field and the quark magnetic moment: the pseudoscalar $D$ ($B$) and vector $D^\ast$ ($B^\ast$) mesons are mixed with each other by a nonzero matrix element~(\ref{eq:spin_mix_1}).
On the other hand, the transverse components of vector mesons are split by the Zeeman effect for the mesons by a nonzero matrix element from Eq.~(\ref{eq:zeeman_1}), which is different from the quarkonia.
This is because the magnetic moments of qurkonia are basically zero, whereas those of heavy-light mesons are nonzero.
Neutral heavy-light mesons such as $D$, $B$, and $B_s$ mesons in a magnetic field were investigated by the constituent quark model \cite{Yoshida:2016xgm}.
The spin mixing effects were also studied by effective Lagrangian~\cite{Gubler:2015qok,Mishra:2020wbo} and should be included in the results estimated from QCD sum rules~\cite{Machado:2013yaa,Gubler:2015qok}.

Furthermore, the heavy-light mesons include a constituent light quark, which is distinct from quarkonia without light quarks.
One of the origins of the the constituent light-quark mass is believed to be the chiral condensate generated from the QCD vacuum.
When we switch on a magnetic field, the value of the chiral condensate increases, which is the so-called magnetic catalysis~\cite{Klevansky:1989vi,Suganuma:1990nn,Klimenko:1990rh,Klimenko:1991he,Klimenko:1992ch,Gusynin:1994re,Gusynin:1994va,Gusynin:1994xp,Gusynin:1995nb}. Hence, the magnetic catalysis can contribute to the masses of heavy-light mesons.
In other words, the mass shifts of heavy-light mesons would be useful for probing the magnetic catalysis.
Such an effect can be investigated by inputting the magnetic-field dependence of the chiral condensate into the OPE side of QCD sum rules \cite{Gubler:2015qok} as well as by artificially changing the constituent quark mass parameter within the constituent quark model \cite{Yoshida:2016xgm}.

\section{Summary and other topics} \label{sec:8}

As other topics which we did not cover in this review, time-dependent phenomena are also important.
In realistic HICs, a magnetic field formed during a collision rapidly damps if the collision energy is large enough.
With the smaller collision energy, the duration time of the magnetic field tends to be longer.
Such a time-dependent magnetic field would lead to various phenomena for quarkonia.
Theoretically, one can consider the time-dependent Schr\"odinger equation and can investigate the time evolution of magnetized quarkonia.
For works focusing on the time evolution of charmonia in a magnetic field, see Refs.~\cite{Guo:2015nsa,Suzuki:2016fof,Dutta:2017pya,Hoelck:2017dby,Bagchi:2018mdi,Iwasaki:2021kms,Iwasaki:phd}.

Furthermore, the heavy quark transport in a magnetic field (and at finite temperature)~\cite{Giataganas:2012zy,Giataganas:2013hwa,Giataganas:2013zaa,Dudal:2014jfa,Fukushima:2015wck,Finazzo:2016mhm,Dudal:2018rki,Kurian:2019nna,Kurian:2020kct,Singh:2020faa} is also interesting.
In such a situation, one can expect an anisotropy of the diffusion coefficients and drag force, which can be estimated by using perturbative calculation~\cite{Fukushima:2015wck,Kurian:2019nna,Kurian:2020kct,Singh:2020faa} and holographic approaches~\cite{Giataganas:2012zy,Giataganas:2013hwa,Giataganas:2013zaa,Dudal:2014jfa,Finazzo:2016mhm,Dudal:2018rki}.
In particular, the directed flows $v_1=\braket{p_x/p_T}$, where $p_x$ and $p_T$ are the momenta along the impact parameter and the reaction plane, respectively, of $D^0$ and $\bar{D}^0$ mesons were observed in STAR at RHIC~\cite{Adam:2019wnk} and ALICE at LHC~\cite{Acharya:2019ijj}, which is much larger than those of light hadrons.
This observation might be an evidence of electromagnetic fields in the initial stage of collisions.
Such an enhancement of $D$ meson flows were predicted by a Langevin approach for heavy quarks with electromagnetic field \cite{Das:2016cwd}, which was motivated by a similar picture for charged pions and protons~\cite{Gursoy:2014aka}.
For other discussions, see Refs.~\cite{Chatterjee:2017ahy,Chatterjee:2018lsx,Oliva:2020mfr,Oliva:2020doe,Sun:2021psy}.

Lastly, we comment on lattice QCD simulations.
By using lattice QCD simulations, one can study situations with a magnetic field much stronger than fields produced in HIC experiments.
Lattice simulations can numerically extract the mass spectra, wave function profiles, and spectral functions of hadrons, so that through such observables, the mass spectra taking into account spin mixing and level repulsions and the deformation of the wave functions can be elucidated (see Refs.~\cite{Bali:2011qj,Luschevskaya:2012xd,Hidaka:2012mz,Luschevskaya:2014lga,Luschevskaya:2015cko,Bali:2017ian,Hattori:2019ijy,Bignell:2020dze,Ding:2020hxw} for charged or neutral pions, Refs.~\cite{Luschevskaya:2012xd,Hidaka:2012mz,Luschevskaya:2014lga,Bali:2017ian,Luschevskaya:2018chr,Hattori:2019ijy} for charged or neutral $\rho$ mesons, and Ref.~\cite{Ding:2020hxw} for strange mesons such as kaons and $\eta_{s\bar{s}}$).
The various properties of magnetized quarkonia, which we have reviewed in this paper, will be examined by lattice simulations implementing a magnetic field.

\begin{acknowledgements}
This work was supported by Japan Society for the Promotion of Science (JSPS) KAKENHI (Grants Nos. JP17K14277, JP19H05159, JP19J13655, JP20K03959, and JP20K14476).
\end{acknowledgements}

\appendix

\section{S-wave charmonia} \label{App:1}
In this Appendix, we briefly review the mass spectra of $S$-wave charmonia and wave function deformation as a characteristic phenomenon of quarkonia in a magnetic field.
In Refs.~\cite{Suzuki:2016kcs,Yoshida:2016xgm}, the authors evaluated the mass spectra and wave functions of quarkonia from the constituent quark model under a constant (or static) magnetic field~\cite{Alford:2013jva} and the cylindrical Gaussian expansion method (CGEM) \cite{Suzuki:2016kcs,Yoshida:2016xgm} which is a numerical approach to solve the anisotropic few-body systems.  

\begin{figure}[t!]
    \centering
    \includegraphics[clip, width=1.0\columnwidth]{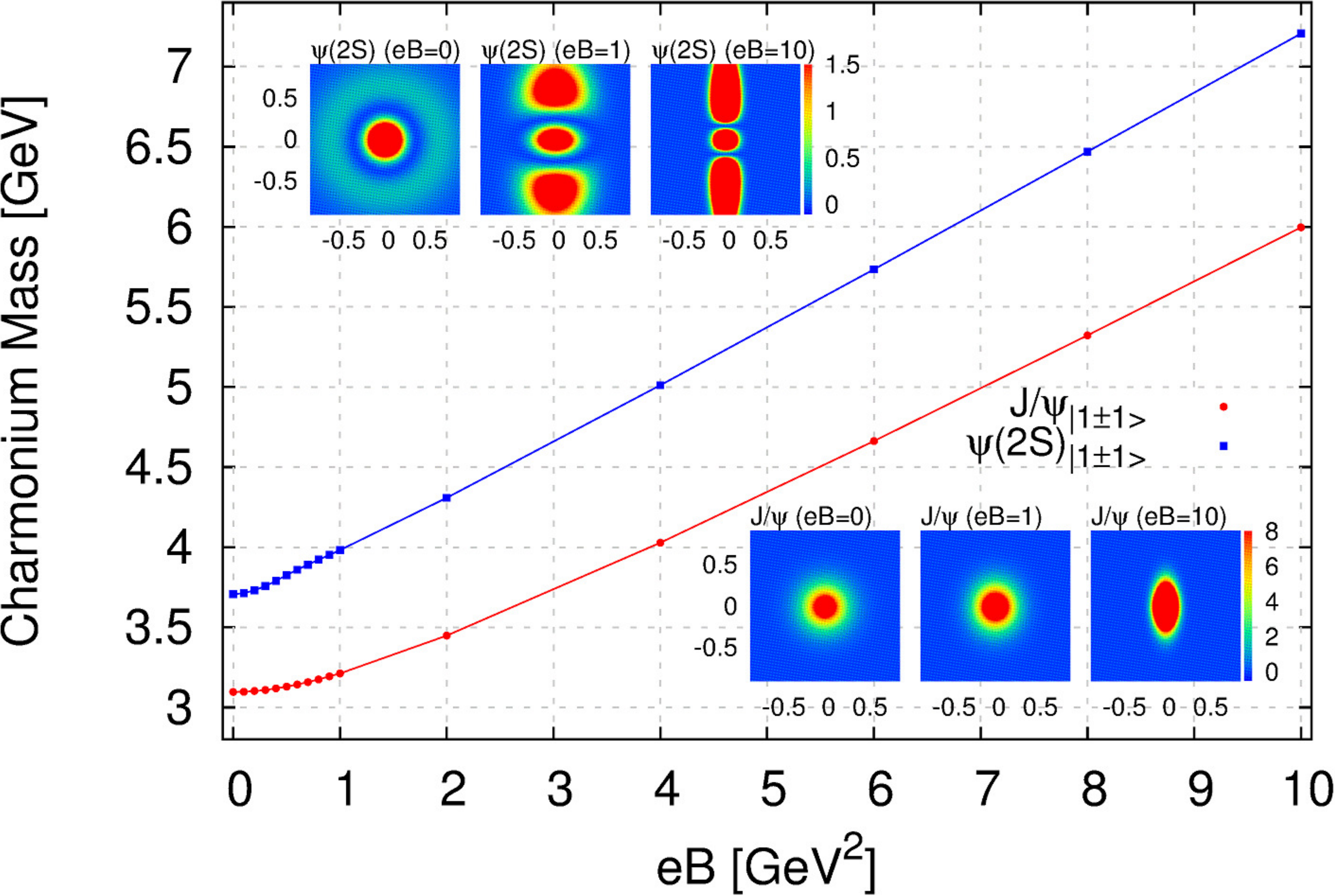}
    \includegraphics[clip, width=1.0\columnwidth]{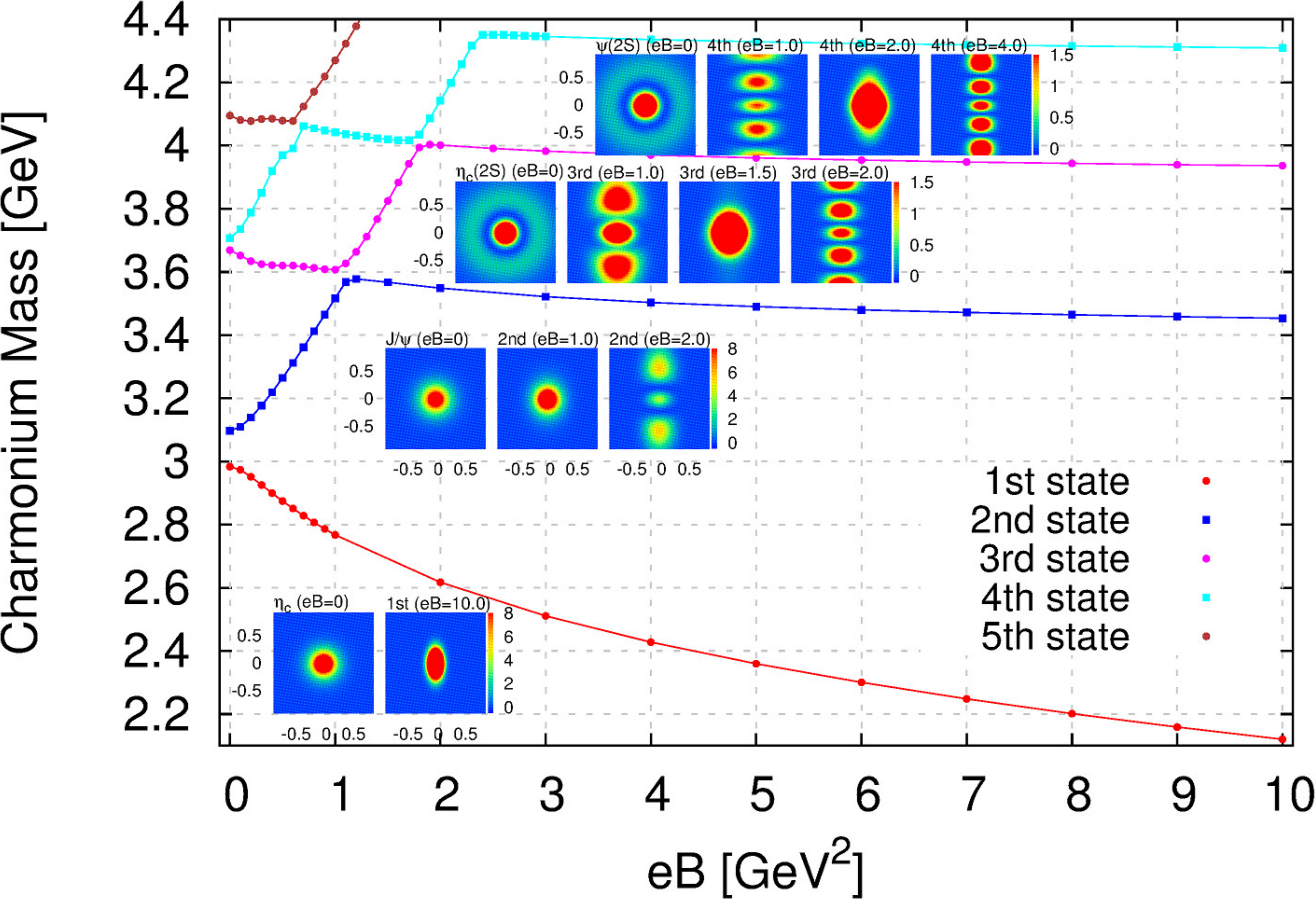}
    \caption{Mass spectra and probability densities of the $S$-wave charmonia in a magnetic field \cite{Yoshida:2016xgm}.
Upper: $S_z=\pm1$ states from $J/\psi$ and $\psi(2S)$.
Lower: $S_z=0$ states from $\eta_c(1S)$, $J/\psi$, $\eta_c(2S)$, and $\psi(2S)$.}
    \label{Fig:swave_c}
\end{figure}

In Fig.~\ref{Fig:swave_c}, we show the mass spectra and probability densities (namely, the squares of wave functions) on the $\rho$-$z$ plane.
The $S$-wave charmonia has the four states below the two-meson ($D\bar{D}$) threshold: the spin-singlet $\eta_c(1S)$-$\eta_c(2S)$ and the spin-triplet $J/\psi$-$\psi(2S)$.
The upper panel of Fig.~\ref{Fig:swave_c} shows the magnetic-field dependence of the masses of the two lowest vector charmonia, $J/\psi$ (red points) and $\psi(2S)$ (blue points), with $S_z=\pm1$.
These states do not mix with different spin eigenstates.
With increasing magnetic field, the masses increase by quark Landau levels (or harmonic-oscillator-type potential in the $\rho$ direction).
At the same time, their wave functions are squeezed on the $\rho$ plane, and then, they are stretched along the $z$ axis, as shown in small windows.
Here, the row of three small windows shows the probability densities at $eB = 0$, $1.0$, and $10.0 \ \mathrm{GeV}^2$.
For example, at $eB = 0$ (the left), we can obtain the spherical (or isotropic) wave functions.
At $eB = 1.0 \ \mathrm{GeV}^2$ (the middle), the $1S$ state is not significantly modified, while the $2S$ state is drastically deformed and its form is the first excitation not in the radial direction but in the $z$ direction.
At $eB = 10.0 \ \mathrm{GeV}^2$ (the right), the $1S$ state becomes a ``cigar-shaped" wave function, and $2S$ state becomes ``rod-shaped".
Note that these deformation is equivalent to mixing between the $S$-wave and higher partial waves with an even number of orbital angular momentum ($L=2,4,\cdots$).
In this sense, radial eigenstates such as $1S$ and $2S$ are no longer ``true" eigenstates, and the true eigenstates in finite magnetic fields are represented as mixing states between different partial waves, which reflects the fact that a magnetic field violates the spherical symmetry, and the orbital angular momentum is no longer a good quantum number.

As another interesting behavior in the upper panel of Fig.~\ref{Fig:swave_c}, one can see a {\it linear increase} of the charmonium masse in the strong-magnetic-field region.
This behavior can be easily understood by the picture of quark Landau levels.
The energies of nonrelativistic Landau levels for a one-body quark with a mass $m$ are represented as $(|qB|/m)(n+1/2)$ with an integer $n$.
In the strong-field limit, this energy is dominated by the lowest Landau level ($n=0$).
By substituting $|qB|=(2/3)|eB|$ and $m_c=1.8$ GeV into this expression, the sum of the mass shifts of the two quarks is estimated to be $0.37 |eB|$ GeV, which is consistent with the slope obtained from the numerical results in Fig.~\ref{Fig:swave_c}.
Thus, the magnetic behavior of $S_z=\pm1$ states in the strong-field limit can be well approximated by a picture with single quarks.
In addition, such a rough estimate seems to be slightly larger than the slope of the $1S$ state.
This tendency may be caused by an attractive mass shift from the Coulomb potential.

The lower panel of Fig.~\ref{Fig:swave_c} shows the mass spectra of $S_z=0$ states, where the spin-singlet eigenstates are mixed with the $S_z=0$ components of the spin-triplet states.
Due to the level repulsion between the two states, the mass of the first state (red points) starting from $\eta_c(1S)$ decreases in nonzero magnetic fields, while that of the second state (blue points) starting from $J/\psi$ increases.
At $eB = 1.0$-$1.1 \ \mathrm{GeV}^2$, the second state approaches the third state (magenta points) from $\eta_c(2S)$.
After the crossing, the second state becomes a $2S$-like wave function, and the third state does $1S$-like.
Similarly, we find the crossing between the third and fourth states at $eB = 1.8$-$1.9 \ \mathrm{GeV}^2$ and the crossing between the fourth and fifth states at $eB = 0.6$-$0.7 \ \mathrm{GeV}^2$.
Thus, the wave functions of excited states are more sensitive to a magnetic field than the ground states.
We emphasize that the position of a crossing point between two states is useful as a visible guideline to determine the mass spectrum under a magnetic field.

\section{$S$-wave bottomonia} \label{App:2}
Next, we compare the magnetic responses of charmonia and bottomonia.
Significant differences between the charm and bottom quarks are (i) the electric charges and (ii) the quark masses.\footnote{Strictly speaking, the detail of the potential, such as the Coulomb potential, should be also different.}  
The absolute value of the electric charge of bottom quarks is $|q_b| = (1/3)|e|$ which is two times smaller than that of charm quarks $|q_c| = (2/3)|e|$.
The masses of bottom quarks are approximately three times heavier than those of charm quarks.
Hence, we can expect that bottomonia are less sensitive to a magnetic field than charmonia.

\begin{figure}[t!]
    \centering
    \includegraphics[clip, width=1.0\columnwidth]{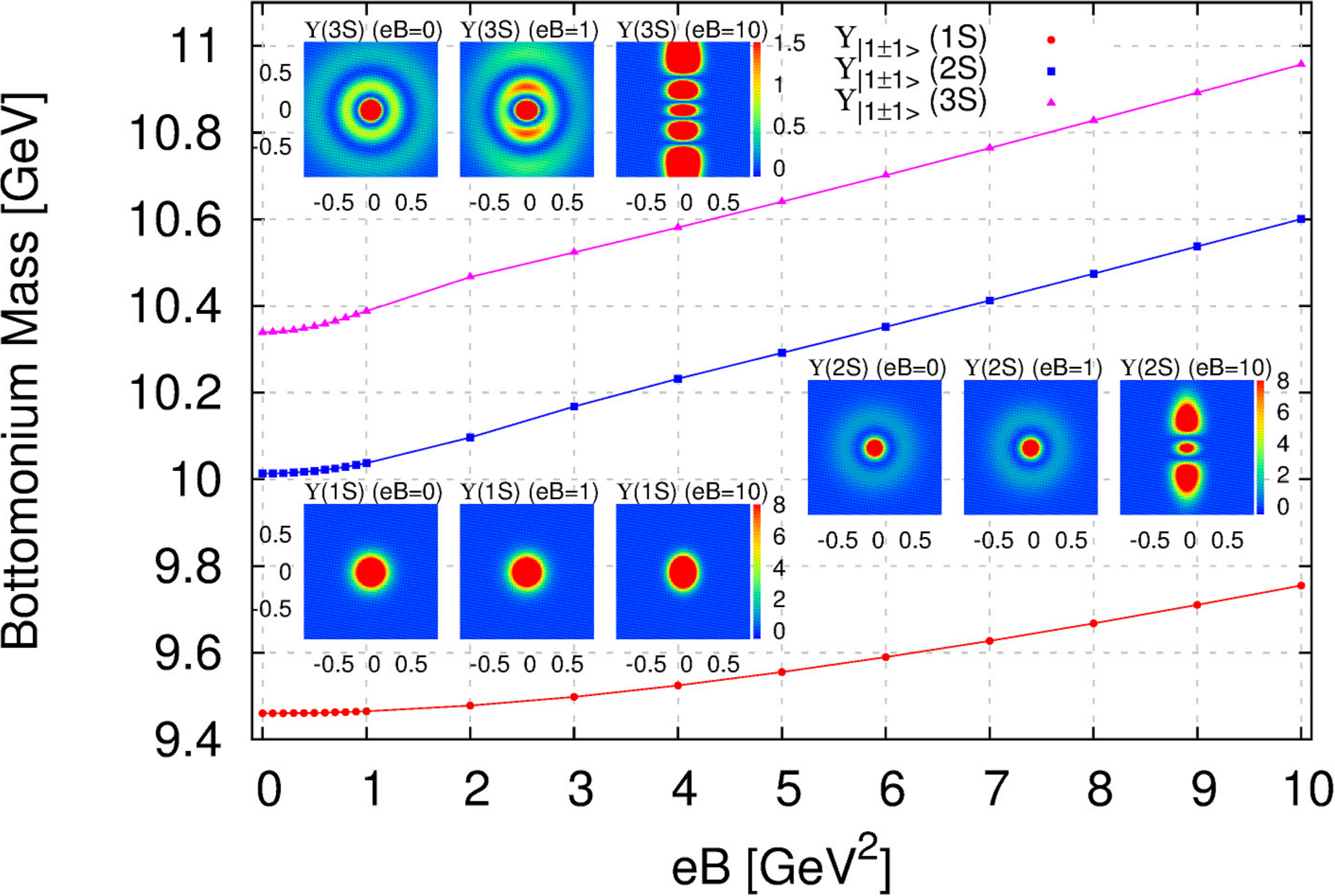}
    \includegraphics[clip, width=1.0\columnwidth]{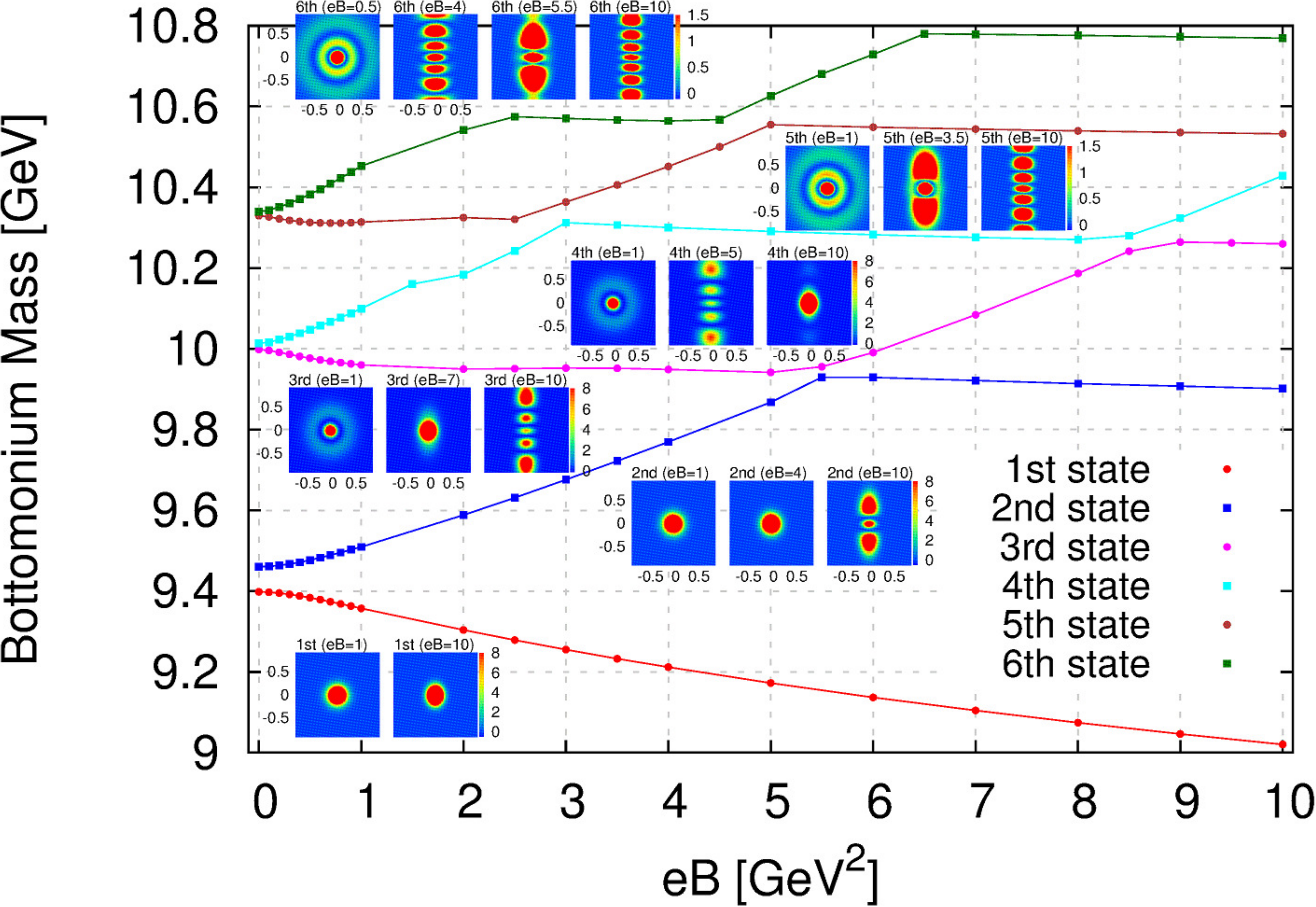}
    \caption{Mass spectra and probability densities of the $S$-wave bottomonia in a magnetic field \cite{Yoshida:2016xgm}.
Upper: $S_z=\pm1$ states from $\Upsilon(1S)$, $\Upsilon(2S)$, and $\Upsilon(3S)$.
Lower: $S_z=0$ states from $\eta_b(1S)$, $\Upsilon(1S)$, $\eta_b(2S)$, $\Upsilon(2S)$, $\eta_b(3S)$, and $\Upsilon(3S)$.}
    \label{Fig:swave_b}
\end{figure}

In Fig.~\ref{Fig:swave_b}, we show the mass spectra of bottmonia, obtained in Ref.~\cite{Yoshida:2016xgm}.
The $S$-wave bottomonia have six states below the two-meson ($B\bar{B}$) threshold: the spin-singlet $\eta_b(1S)$-$\eta_b(2S)$-$\eta_b(3S)$ and spin-triplet $\Upsilon(1S)$-$\Upsilon(2S)$-$\Upsilon(3S)$.
The upper panel of Fig.~\ref{Fig:swave_b} shows the spectra of $S_z=\pm1$ states.
At $eB = 1.0 \ \mathrm{GeV}^2$, the wave functions of $1S$, $2S$, and $3S$ states are almost spherical, which is a situation different from the charmonum spectrum, where $\psi(2S)$ is significantly deformed.
Even at $eB = 10.0 \ \mathrm{GeV}^2$, the wave function of the ground state is still spherical.
The insensitivity of bottomonia to magnetic fields can be seen also in the magnitude of the mass shift.

The lower panel of Fig.~\ref{Fig:swave_b} shows the mass spectra of $S_z=0$ states.
With the same mechanism as the charmonia, one can see the level crossing between different levels.
The crossing point between the second and third states is at $eB = 5.5 \ \mathrm{GeV}^2$, while the corresponding point in the charmonium spectrum is $eB = 1.0$-$1.1 \ \mathrm{GeV}^2$.
Similarly, one can see that $eB = 8.5 \ \mathrm{GeV}^2$ for the third and fourth states, $eB = 3.0 \ \mathrm{GeV}^2$ for the fourth and fifth states, and $eB = 5.0 \ \mathrm{GeV}^2$ for the fifth and sixth states.

\section{$P$-wave charmonia and hadronic Paschen-Back effect} \label{App:3}
In this Appendix, we briefly review the {\it hadronic Paschen-Back effect} (HPBE)~\cite{Iwasaki:2018pby} which is analogous to a phenomenon observed in atomic physics, the Paschen-Back effect~\cite{Paschen:1921}.
The HPBE is realized by interplay between an orbital angular momentum of hadrons and a magnetic field.
The HPBE occurs in various hadrons with a finite orbital angular momentum, but here we focus on the $P$-wave charmonia: the spin-singlet $h_c$ ($^1 \! P_1$) and spin-triplet $\chi_{c0}$ ($^3 \! P_0$), $\chi_{c1}$  ($^3 \! P_1$), and $\chi_{c2}$ ($^3 \! P_2$), where we used a notation $^{2S+1} \! P_J$ with the total angular momentum $J=L+S$, orbital angular momentum $L$, and spin angular momentum $S$.
In zero magnetic field, $L_z$ and $S_z$ are not conserved by the existence of the LS and tensor coupling, and the good quantum numbers are $J$, $L$, and $S$.

When a magnetic field is switched on, only the $z$ component of the total angular momentum, $J_z=L_z+ S_z$, is a conserved quantity.
In addition, if the magnetic field becomes stronger than the scale of the spin-orbit splitting, $L_z$ and $S_z$ are approximately good quantum numbers.
Then, instead of the $\ket{J;LS}$ bases used in weak fields, it may be convenient to use a new basis $\ket{J_z;L_zS_z}$ in a strong field.
We call such a strong magnetic-field region the {\it PB region} (or the {\it PB limit} for the strong-field limit).\footnote{Note that even in the PB limit, the tensor coupling exists and mixes $L_z$ and $S_z$, so that these are not conserved.}
In particular, the wave functions of $P$-wave charmonia in the PB region are approximately expressed as 
\begin{align}
\Psi_{L_z;S_{1z}S_{2z}}(\rho,z,\phi) = \Phi_{L_z}(\rho, z) Y_{1 L_z}(\theta,\phi)\chi(S_{1z}, S_{2z}), \label{eq.PBconf}
\end{align}
where $(\rho,z,\phi)$ is the cylindrical coordinate, and $\tan\theta=\rho/z$.
This configuration is composed of two parts: (i) the spatial part and (ii) spin part.
The spatial wave function is defined by $\Phi(\rho,z)$ and the spherical harmonics $Y_{1L_z}(\theta,\phi)$.\footnote{Note that this configuration contains partial waves such as $L=1,3,5,\cdots$.
Nevertheless, the factorization of Eq.~(\ref{eq.PBconf}) is valid because all the partial waves with a same $L_z$ always share the same factor of $e^{\pm i\phi}\sin\theta\propto Y_{1\pm1}$ or $\cos\theta\propto Y_{10}$.
}
The spin wave function is defined as $\chi(S_{1z}, S_{2z})$, where $S_{1z}$ and $ S_{2z}$ are the third components of the spin of the charm quarks.

\begin{figure}[t!]
    \centering
    \includegraphics[clip, width=1.0\columnwidth]{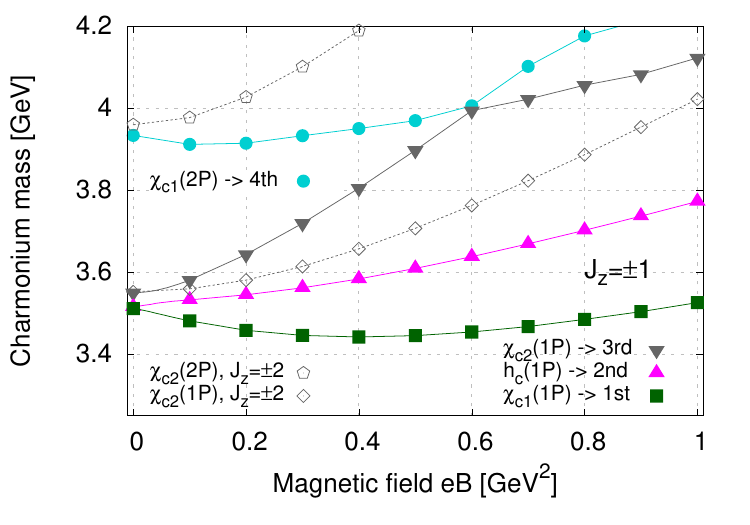}
    \includegraphics[clip, width=1.0\columnwidth]{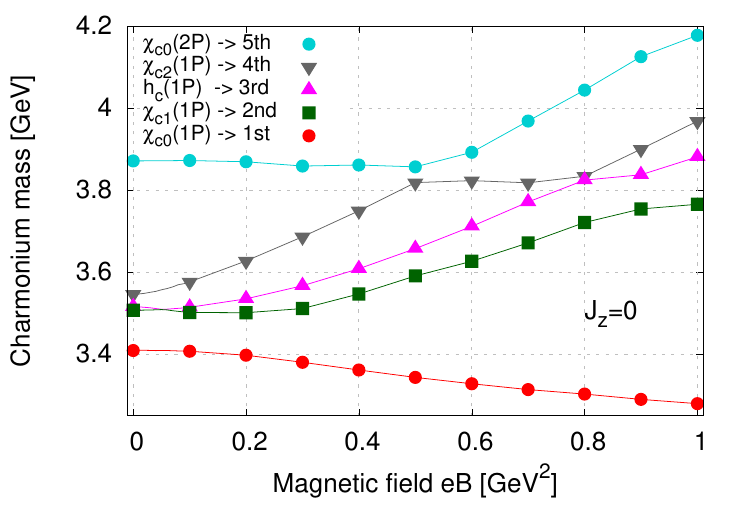}
    \caption{Mass spectra of the $P$-wave charmonia in a magnetic field \cite{Iwasaki:2018pby}.
Upper: $J_z=\pm2$ states from $\chi_{c2}(1P)$ and $\chi_{c2}(2P)$ and $J_z=\pm1$ states from $\chi_{c1}(1P)$, $h_c(1P)$, $\chi_{c2}(1P)$, and $\chi_{c1}(2P)$.
Lower: $J_z=0$ states from $\chi_{c0}(1P)$, $\chi_{c1}(1P)$, $h_c(1P)$, $\chi_{c2}(1P)$, and $\chi_{c0}(2P)$.}
    \label{Fig_mass}
\end{figure}

The mass spectra of $P$-wave charmonia in a magnetic field is shown in Fig.~\ref{Fig_mass}.
The upper panel shows both the $J_z = \pm2$ and $J_z = \pm1$ states, and the lower panel shows $J_z = 0$.
The $J_z = \pm2$ states starting from $\chi_{c2}$ are not mixed with other states, so that their masses continue to increase, which is similar to the $S_z=\pm 1$ states of the $S$-waves.
The $J_z = \pm1,0$ states exhibit the mixing between spin eigenstates, which is similar to the $S_z=0$ states of the $S$-waves.
As a result of the mixing and level repulsion, the masses of the lowest states decrease as the magnetic field increases, while higher states becomes heavier.
We find the level crossings between the $1P$-like state and $2P$-like state at $eB = 0.6 \ \mathrm{GeV}^2$ for $J_z=\pm1$, and at $eB = 0.5$ and $0.8\ \mathrm{GeV}^2$ for $J_z=0$.
Thus, the magnetic field for the crossing between $1P$ and $2P$ states is smaller than that for the corresponding $1S$ and $2S$ states.

\begin{figure}[t!]
    \centering
    \includegraphics[clip, width=1.0\columnwidth]{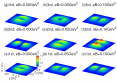}
    \caption{Probability densities of wave functions of $P$-wave charmonia for $J_z=\pm1$ in magnetic fields \cite{Iwasaki:2018pby}.
The vertical axis is $|\Psi(\rho, z, \phi)|^2$, and the horizontal plane is represented by the $\rho$ and $z$ axes, where $\rho$ ($z$) is the spatial direction perpendicular (parallel) to the magnetic field.
    }
    \label{Fig_WF}
\end{figure}

In Fig.~\ref{Fig_WF}, we show the probability densities on the $\rho$-$z$ plane, defined as $|\Psi(\rho, z, \phi)|^2$ for the $J_z=\pm1$ states.
At $eB=0$ [(a), (d), and (g)], the ``1st", ``2nd", and ``3rd" wave functions correspond to $\chi_{c1}, h_c$, and $\chi_{c2}$ states, respectively.
With increasing magnetic field, the wave functions are gradually deformed.
At $eB=0.1 \ \mathrm{GeV}^2$ [(c), (f), and (i)], one can see probability densities characterized by $|L_z|$.
Note that the Hamiltonian has inversion symmetry along the $z$-axis, so that we see the same physics for both the $L_z=\pm 1$ states.
For example, the basis function with $L_z=\pm1$ is characterized by the factor of $r Y_{1\pm1}(\theta, \phi)\propto r\cos\theta e^{\pm i\phi}=\rho e^{\pm i\phi}$.
This factor means that the magnitude of the wave function on the $z$ axis is zero [see (c) and (i)].
On the other hand, the basis function with $L_z=0$ is based on the factor of $r Y_{10}(\theta, \phi) \propto r\sin\theta=z$.
This means that the wave function amplitude on the $\rho$ axis is zero [see (f)].
Thus, in the PB limit, the spatial wave function of an eigenstate is purified to either wave function with $L_z=\pm1$ or $L_z=0$.
The magnetic field of $eB=0.1 \ \mathrm{GeV}^2$ can be well approximated as the PB limit, and thus one can observe drastic deformation by the HPBE even in small magnetic fields.

We emphasize that the deformation mechanism by the HPBE is quite different from that in $S$-wave quarkonia.
The deformation in $S$-waves is induced by quark Landau levels, or equivalently the mixing between an $S$-wave and higher partial waves with an even orbital angular momentum, $L=2,4,\cdots$.
On the other hand, the origin of the HPBE is ``resolving" the mixed states originating from the LS coupling between two quarks.
Therefore, the scale of magnetic fields that induces the HPBE corresponds to the LS splitting of hadron masses, which leads to drastic deformation in relatively smaller magnetic field than that for $S$-waves.

\bigskip
\bigskip

\bibliography{quarkonium-Bfield}
\end{document}